\documentclass[a4paper,fleqn]{cas-dc}
\RenewDocumentCommand \printorcid { } { }
\usepackage[numbers]{natbib}
\usepackage[normalem]{ulem}
\usepackage{amsmath,amssymb,amsfonts}
\usepackage{graphicx}
\usepackage{textcomp}
\usepackage{xcolor}
\usepackage{url}

\newcommand{\jj}{\mathrm{j}}

\begin{document}
\let\WriteBookmarks\relax
\def\floatpagepagefraction{1}
\def\textpagefraction{.001}

\shorttitle{Split Coaxial Cable Medium for Tunable Artificial Dielectrics and Plasmas}
\shortauthors{Zhuravlev et al.}

\title[mode=title]{Split Coaxial Cable Medium for Tunable Artificial Dielectrics and Plasmas}

\author[1]{Alexander Zhuravlev}\fnmark[1]

\author[1]{Jim A. Enriquez}\fnmark[1]\cormark[1]
\ead{jim.enriquez@metalab.ifmo.ru}

\author[1,2]{Pavel A. Belov}

\author[3]{Juan D. Baena}

\affiliation[1]{organization={School of Physics and Engineering, ITMO University},
            city={Saint Petersburg},
            postcode={191002},
            country={Russia}}

\affiliation[2]{organization={School of Engineering, New Uzbekistan University},
            city={Tashkent},
            postcode={100007},
            country={Uzbekistan}}

\affiliation[3]{organization={Department of Physics, Universidad Nacional de Colombia},
            city={Bogot\'a},
            postcode={111321},
            country={Colombia}}

\cortext[1]{Corresponding author.}
\fntext[1]{Alexander Zhuravlev and Jim A. Enriquez contributed equally to this work.}

\begin{abstract}
We introduce the Split Coaxial Cable Medium (SCCM), a mechanically tunable, capacitively loaded wire medium supporting artificial-dielectric and artificial-plasma Bloch regimes with high in-plane isotropy. Its unit cell comprises coaxial conductors interrupted by axial gaps. Relative axial displacement continuously varies their capacitive overlap and series capacitance while preserving the transverse lattice geometry and leaving the unit-cell inductance approximately unchanged. Equivalent RLC parameters are derived directly from the geometry and incorporated into complementary analytical models. A spatially dispersive local-field model predicts the Bloch dispersion and isofrequency contours, whereas a multilayer homogenization model provides the modal impedance, closed-form estimates of the low-frequency Bloch refractive index and plasma frequency, and loss-inclusive finite-slab scattering. Full-wave eigenmode and finite-slab simulations validate the predictions. Over the investigated displacement range, simulations yield tunabilities of $46\%$ in the low-frequency Bloch refractive index, from $n_{0,\min}=1.42$ to $n_{0,\max}=2.28$, and $21\%$ in the plasma frequency, from $f_{\mathrm{p},\min}=9.45~\mathrm{GHz}$ to $f_{\mathrm{p},\max}=11.61~\mathrm{GHz}$, while confirming high in-plane isotropy near the $\Gamma$ point in both regimes. These characteristics make the SCCM promising for gradient-index devices, directive antennas, and tunable plasma haloscopes.
\end{abstract}

\begin{highlights}
\item A tunable split coaxial medium supports dielectric-like and plasma-like Bloch bands.
\item Axial displacement tunes capacitance while leaving inductance nearly unchanged.
\item Simulations show 46\% refractive-index and 21\% plasma-frequency tunability.
\item Both passbands exhibit nearly isotropic in-plane propagation near the $\Gamma$ point.
\end{highlights}


\maketitle

\section{Introduction}
Metamaterials enable engineered electromagnetic responses beyond those available in natural materials, providing advanced control over electromagnetic fields~\cite{Alu2025}. In particular, engineered permittivity responses have been exploited to enhance antenna directivity~\cite{lovat2007combinations}, improve radiation efficiency and impedance matching through high-permittivity substrates~\cite{syed2013front}, and increase operational bandwidth using epsilon-near-zero materials~\cite{jafargholi2024enabling,boas2025epsilon}. Wire media (WM) constitute an established platform for realizing artificial-plasma responses~\cite{BelovT2002,simovski2004}. When capacitively loaded, WM can additionally support an artificial-dielectric regime~\cite{BelovS2002}.

\begin{figure}
\includegraphics[width=0.9\columnwidth]{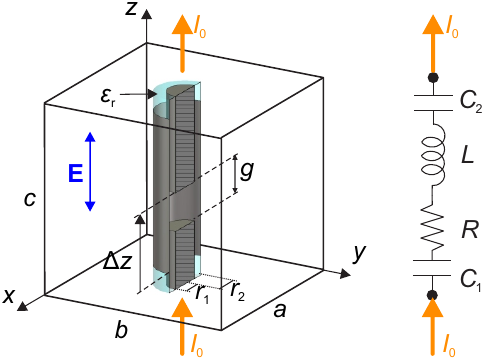}
  \centering
  \caption{Unit cell of the SCCM and its equivalent RLC circuit. The average electric field $\mathbf{E}$ is oriented along the $z$ direction, and the reference current $I_0$ flows along the same direction. The space between the two conducting cylinders is filled with a dielectric (light blue) and both cylinders are interrupted by axial gaps of width $g$. The inner conductor is solid with radius $r_1$, whereas the outer conductor has inner radius $r_2$ and wall thickness $t_{\mathrm{m}}$. The fixed unit-cell parameters are $a=b=c=10~\mathrm{mm}$, $r_1=0.3~\mathrm{mm}$, $r_2=0.5~\mathrm{mm}$, $t_{\mathrm{m}}=18~\mu\mathrm{m}$, $g=2~\mathrm{mm}$, and $\varepsilon_r=2$, whereas the relative axial displacement $\Delta z$ is varied for tuning.}
  \label{Fig1-unitCell}
\end{figure}

Plasma-like metamaterials have attracted considerable interest in
fundamental physics, including axion and dark-photon detection~\cite{gelmini2020,millar2023alpha,Lawson2019} and
gravitational-wave detection~\cite{capdevilla2025}. In WM-filled cavities, the plasma-like response is associated with long-wavelength modes near an epsilon-near-zero condition~\cite{balafendiev2022resonator, enriquez2025uniformfieldmicrowavecavities}. Several of these applications require a tunable plasma frequency~\cite{millar2023alpha}. Existing mechanically tunable WM typically vary the transverse wire spacing~\cite{Kowitt2023,sakhno2025volume,Lindahl2026}, whereas a related approach employs rotating obround meta-atoms~\cite{balafendiev2025tunableepsilonnearzero}. These mechanisms require transverse displacement or reorientation of the constituent elements, demanding lateral clearance and potentially complicating mechanical actuation in enclosed conducting cavities. This motivates alternative tuning strategies based on axial motion.

Tunable artificial dielectrics, in turn, provide control over the low-frequency refractive index and have been investigated for gradient-index (GRIN) devices such as microwave lenses~\cite{Chen2023}. Many existing implementations employ anisotropic unit cells~\cite{Wan2014,Lin2018}, potentially restricting direction-independent operation, bandwidth, or tuning range. Together, these considerations motivate a mechanically simple WM capable of tuning both the low-frequency refractive index and the plasma frequency while preserving the transverse lattice geometry and high in-plane isotropy.

A closely related antecedent is the mechanically tunable sliding split-strip metamaterial reported in the preliminary study of Ref.~\cite{Enriquez2025p}. That study demonstrated tunable dispersion in a noncoaxial split-strip geometry but focused primarily on the dispersion curves. It did not derive the equivalent-circuit parameters directly from the geometrical dimensions or investigate material losses, finite-slab scattering, and electromagnetic-field and current distributions. It therefore served as an initial proof of concept, whereas the present work provides a more comprehensive analytical and physical description.

Building on this concept, we introduce the Split Coaxial Cable Medium (SCCM). The coaxial topology produces an approximately azimuthally invariant surface-current distribution, concentrates the electric field primarily within the capacitive overlap regions, and allows the magnetic field to extend outside the outer conductor. We derive the equivalent RLC parameters directly from the SCCM geometry and incorporate them into complementary local-field and multilayer homogenization models. Together, these models provide the Bloch dispersion, isofrequency contours, modal impedance, closed-form estimates of the low-frequency Bloch refractive index and plasma frequency, and loss-inclusive finite-slab scattering. The relative axial displacement between the inner and outer conductors provides continuous tuning of the low-frequency Bloch refractive index and plasma frequency while preserving the transverse lattice geometry. Full-wave simulations yield tunabilities of $46\%$ in the low-frequency Bloch refractive index and $21\%$ in the plasma frequency while confirming high in-plane isotropy, making the SCCM a promising platform for GRIN devices and mechanically tunable plasma haloscopes.

\section{Theoretical Analysis and Tunability Mechanism}\label{section2:theory}

\subsection{Equivalent RLC circuit}\label{subsec_circuit_model}

\begin{figure}
  \includegraphics[width=0.9\columnwidth]{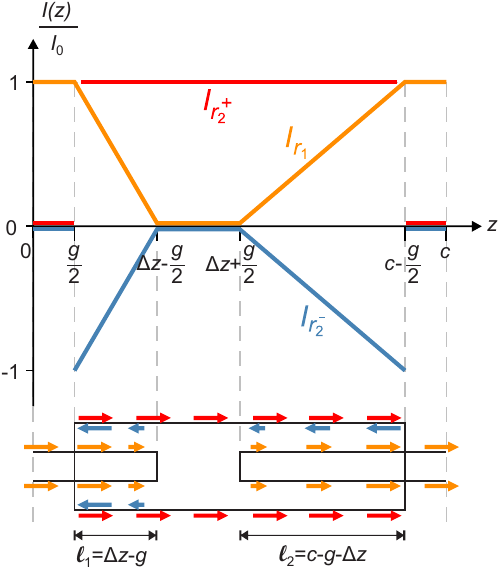}
  \centering
  \caption{Piecewise linear approximation of the current distribution. The electrical current spreads over three different surfaces and can be separated as: $I_{r_1}$ on the inner cylinder (orange line), $I_{r_2^-}$ on the inner face of the outer cylinder (blue line), and $I_{r_2^+}$ on the outer face of the same outer cylinder (red line). A simplified sketch of currents on the structure is shown at the bottom, where the lengths of current arrows represent their magnitudes.}
  \label{Fig2-currentApprox}
\end{figure}

Fig.~\ref{Fig1-unitCell} shows the SCCM unit cell and its corresponding equivalent series-RLC circuit, comprising the resistance $R$, inductance $L$, and capacitances $C_1$ and $C_2$. Axial gaps of width $g$ interrupt the direct conduction paths in both conductors, causing charge to accumulate near the gap edges. The resulting displacement currents across the dielectric couple the two conductors and are represented by the capacitances $C_1$ and $C_2$, which correspond to coaxial overlap regions of axial lengths $l_1$ and $l_2$, respectively. A relative axial displacement $\Delta z$ changes these overlap lengths while preserving $l_1+l_2=c-2g$, thereby tuning the equivalent series capacitance and, consequently, the Bloch response of the SCCM.

To estimate the circuit parameters, we assume that the axial surface-current density is approximately invariant with respect to the azimuthal coordinate. The current on a cylindrical surface of radius $r$ can then be written as
$I_r(z)=\int_0^{2\pi}K_z(r,\phi,z)r\,d\phi\simeq2\pi rK_z(r,z)$.
For an electrically small unit cell, the currents are approximated by the piecewise-linear profiles shown in Fig.~\ref{Fig2-currentApprox}. The inner-conductor current $I_{r_1}$ is approximately constant over the portions aligned with the outer-conductor gap and varies linearly across the two overlap regions. The current $I_{r_2^-}$ on the inner surface of the outer conductor flows in the opposite direction, suppressing the magnetic field inside a conductor whose thickness exceeds the skin depth. By contrast, the exterior current $I_{r_2^+}$ remains approximately constant between the edges of the outer-conductor gap because negligible charge accumulates on that surface. At the gap edges, $I_{r_2^-}$ and $I_{r_2^+}$ have equal magnitudes and opposite directions as the current turns between the two surfaces. The signed currents satisfy
\begin{equation}
I_{r_1}+I_{r_2^-}+I_{r_2^+}=I_0,
\end{equation}
expressing the continuity of the total axial current, including its transfer between conductors through displacement currents.

For a good conductor with conductivity $\sigma$ and skin depth
$\delta=\sqrt{2/(\mu_0\sigma\omega)}$, the time-average surface loss density is $|\mathbf{K}|^2/(2\sigma\delta)$~\cite[Sec.~8.1]{Jackson1999}. Using the current profiles of Fig.~\ref{Fig2-currentApprox}, the equivalent resistance becomes
\begin{equation}
\begin{aligned}
R
&= \frac{2\langle P_{\mathrm{loss}}\rangle}
        {|I_0|^2}                                                   \\
&= \frac{1}{\sigma\delta |I_0|^2}
   \int_0^c
   \Bigg(
   \frac{|I_{r_1}|^2}{2\pi r_1}
   +\frac{|I_{r_2^-}|^2}{2\pi r_2}            
   +\frac{|I_{r_2^+}|^2}{2\pi r_2}
   \Bigg)\,dz                                                        \\
&\approx
\frac{1}{2\pi\sigma\delta}
\left(
\frac{c+g}{3r_1}
+\frac{4c-5g}{3r_2}
\right).
\end{aligned}
\label{eq:resistance}
\end{equation}

The inductance follows from the time-average magnetic energy,
$\langle W_{\mathrm m}\rangle=L|I_0|^2/4$, and is separated into exterior and interior contributions, $L=L^{\mathrm{out}}+L^{\mathrm{in}}$. Outside the outer conductor, the SCCM is approximated as a thin-wire lattice, giving
$L^{\mathrm{out}}\simeq\mu_0c\ln[b/(2\pi r_2)]/(2\pi)$~\cite{Marcuvitz1951}. Within the coaxial region, Ampere's law gives $H_\phi=I_{r_1}/(2\pi r)$. Integrating the corresponding magnetic energy using the piecewise-linear profile of $I_{r_1}$ yields
\begin{equation}
\begin{aligned}
L
&\approx L^{\mathrm{out}}+L^{\mathrm{in}}                                  \\
&\approx
\frac{\mu_0c}{2\pi}
\ln\left(\frac{b}{2\pi r_2}\right)+
\frac{\mu_0(c+g)}{6\pi}
\ln\left(\frac{r_2}{r_1}\right).
\end{aligned}
\label{eq:inductance}
\end{equation}

Within this approximation, $R$ and $L$ are independent of $\Delta z$ because the combined overlap length $l_1+l_2=c-2g$ remains constant. By contrast, the individual capacitances vary strongly with displacement. Approximating each overlap region as a cylindrical capacitor and introducing a fringing-field extension $r_2-r_1$ at each end gives
\begin{equation}
C_i
\approx
\frac{2\pi\varepsilon_0\varepsilon_r}
{\ln(r_2/r_1)}
\left[l_i+2(r_2-r_1)\right],
\qquad i=1,2,
\label{eq}
\end{equation}
where $l_1=\Delta z-g$ and $l_2=c-g-\Delta z$. The total series capacitance is
$C^{-1}=C_1^{-1}+C_2^{-1}$ and therefore varies with $\Delta z$, reaching its maximum in the symmetric configuration $\Delta z=c/2$, for which $C_1=C_2$.
\subsection{Bloch dispersion and modal impedance}\label{Section-Dispersion_relations}
\begin{figure}
 \includegraphics[width=\columnwidth]{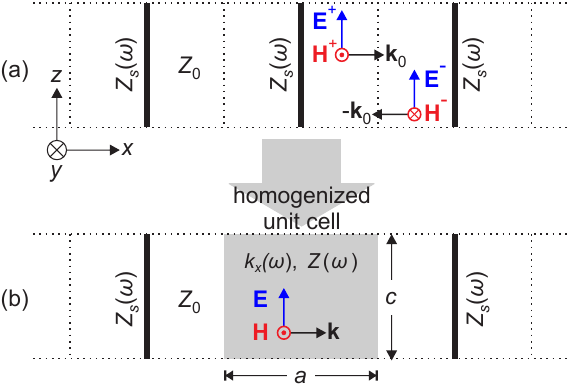}
 \caption{Multilayer homogenization model. (a) The SCCM is represented as a periodic stack of impedance sheets characterized by
$Z_s(\omega)=(b/c)[R+\jj\omega L+(\jj\omega C)^{-1}]$, where $b$ and $c$ are the periods along $y$ and $z$, respectively, and the quantity in brackets is the equivalent series-RLC impedance of the unit cell in Fig.~\ref{Fig1-unitCell}. (b) One period of the stack is replaced by an equivalent homogeneous layer characterized by the Bloch propagation constant $k_x$ and modal impedance $Z$.}
 \label{Fig3-multilayerHomogenization}
\end{figure}

Two complementary analytical models are used in this work. The local-field model retains the higher-order spatial harmonics of the wire lattice and is used to calculate the dispersion relations and isofrequency contours. The multilayer homogenization model replaces each transverse layer with an equivalent impedance sheet. This model is restricted to propagation normal to the sheets and becomes less accurate when the unit cell is not deeply subwavelength. However, it provides compact expressions for the Bloch propagation constant and modal impedance, enabling analytical estimates of the low-frequency refractive index and plasma frequency, as well as finite-slab scattering calculations. Both models use the equivalent RLC parameters derived in Section~\ref{subsec_circuit_model}.

\paragraph{Local-field model}

The local-field approach of Refs.~\cite{BelovS2002,BelovT2002} gives the
Bloch dispersion relation of a capacitively loaded wire medium as
\begin{equation}
    \cos(k_x a)
    =
    \cos\left(k_x^{(0)}a\right)
    +
    \jj\frac{Z_0}{2Z_s}
    \sin\left(k_x^{(0)}a\right),
    \label{eq:dispersion_loadedWM}
\end{equation}
where $k_x$ is the $x$ component of the Bloch wave vector,
$k_0=\omega/c_0$, $Z_0$ is the free-space wave impedance, and $Z_s$ is the spatially dispersive surface impedance of the capacitively loaded wire lattice:
\begin{equation}
\begin{split}
    Z_s
    ={}&
    \jj\frac{Z_0}{2}
    \frac{b k_x^{(0)}}{\pi}
    \ln\left(\frac{b}{2\pi r_2}\right)
    \\
    &+
    \jj\frac{Z_0}{2}k_x^{(0)}
    \sum_{n\neq0}
    \left[
    \begin{aligned}
        &
        \frac{1}{k_x^{(n)}}
        \frac{\sin\left(k_x^{(n)}a\right)}
        {\cos\left(k_x^{(n)}a\right)-\cos(k_xa)}
        \\
        &
        -
        \frac{b}{2\pi|n|}
    \end{aligned}
    \right]
    \\
    &+
    \frac{k_x^{(0)}}{k_0}\frac{b}{c}
    \left(
        R+\jj\omega L^{\mathrm{in}}
        +\frac{1}{\jj\omega C}
    \right).
    \label{eq:surfaceImpedanceSpatiallyDispersive}
\end{split}
\end{equation}
The first two terms in Eq.~\eqref{eq:surfaceImpedanceSpatiallyDispersive} describe the unloaded wire lattice~\cite{BelovS2002}, with the summation accounting for the higher-order Floquet harmonics. The last term incorporates the resistance, internal inductance, and series capacitance of the split coaxial conductors. Only $L^{\mathrm{in}}$ appears in this loading term because the external inductive contribution is already contained in the first term.

The $x$ component of the Bloch wave vector of the $n$th spatial harmonic is
\begin{equation}
    k_x^{(n)}
    =
    -\jj
    \sqrt{
        \left(
            k_y+\frac{2\pi n}{b}
        \right)^2
        +k_z^2-k_0^2
    },
    \label{eq:spatialHarmonics}
\end{equation}
where the square-root branch is selected to ensure the decay of evanescent harmonics. Equations~\eqref{eq:dispersion_loadedWM}--%
\eqref{eq:spatialHarmonics} determine the dispersion relations and isofrequency contours of the SCCM. For propagation normal to the equivalent sheets, $k_y=k_z=0$ and $k_x^{(0)}=k_0$.
Equation~\eqref{eq:dispersion_loadedWM} then has the same functional form as the multilayer dispersion relation below, while $Z_s$ retains the higher-order harmonic contributions.

\paragraph{Multilayer homogenization model}

The multilayer homogenization method of Ref.~\cite{Parra2025} relates two electromagnetically equivalent configurations: (i) a periodic stack of impedance sheets representing the transverse SCCM layers [Fig.~\ref{Fig3-multilayerHomogenization}(a)] and (ii) a homogeneous medium characterized by the Bloch propagation constant $k_x$ and modal impedance $Z$ [Fig.~\ref{Fig3-multilayerHomogenization}(b)]. Equating their transfer matrices over one period gives the Bloch dispersion relation reported in Ref.~\cite[Eq.~(4)]{Parra2025}:
\begin{equation}
    \cos(k_xa)
    =
    \cos(k_0a)
    +
    \jj\frac{Z_0}{2Z_s}
    \sin(k_0a).
    \label{eq:disRel}
\end{equation}
The corresponding normalized modal admittance is
\cite[Eq.~(5)]{Parra2025}
\begin{equation}
    \frac{Z_0}{Z}
    =
    \sqrt{
        1-
        \frac{2}{
            1
            -\jj\frac{2Z_s}{Z_0}\sin(k_0a)
            -\cos(k_0a)
        }
    }.
    \label{eq:admittance}
\end{equation}
The physically appropriate branches of $k_x$ and $Z$ are selected by continuity with the corresponding passband and, in the presence of losses, by passivity.

Within this model, the higher-order lattice-harmonic contributions are
neglected, and the sheet impedance is approximated using the equivalent
circuit of Section~\ref{subsec_circuit_model}:
\begin{equation}
    Z_s
    \approx
    \frac{b}{c}
    \left[
        \jj\omega
        \frac{\mu_0c}{2\pi}
        \ln\left(\frac{b}{2\pi r_2}\right)
        +R
        +\jj\omega L^{\mathrm{in}}
        +\frac{1}{\jj\omega C}
    \right].
    \label{eq:surfImpedance}
\end{equation}
The first term inside the brackets is the external inductive contribution, $L^{\mathrm{out}}$. For normal propagation, $k_x^{(0)}=k_0$ and $Z_0k_0=\omega\mu_0$; therefore, the first term of
Eq.~\eqref{eq:surfaceImpedanceSpatiallyDispersive} reduces to the
corresponding external inductive term in Eq.~\eqref{eq:surfImpedance}. The remaining terms account for $R$, $L^{\mathrm{in}}$, and $C$, while the factor $b/c$ converts the unit-cell impedance into the equivalent sheet impedance.

\paragraph{Validation of the dispersion models}
\begin{figure}
  \centering
  \includegraphics[width=0.9\columnwidth]{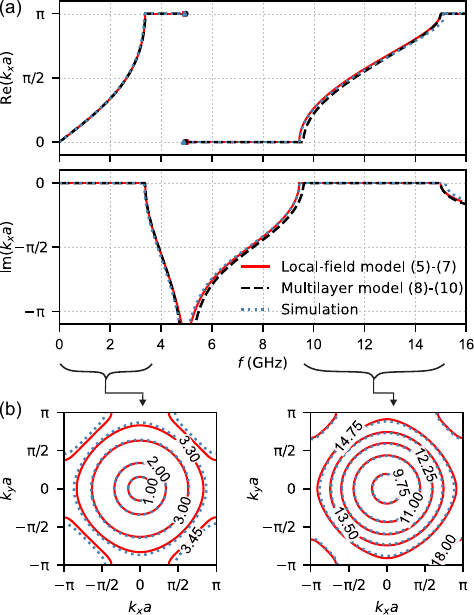}
  \caption{Theoretical and simulated dispersion characteristics of the SCCM shown in Fig.~\ref{Fig1-unitCell} with $\Delta z=0.5c$: (a) dispersion relation for propagation in the $x$-direction and (b) isofrequency contours for arbitrary propagation directions in the $xy$-plane, corresponding to the first (dielectric-type) and second (plasma-type) passbands. The numbers adjacent to the isofrequency contours indicate frequency in
GHz.} 
  \label{Fig4-dispersionRelation}
\end{figure}
We validated the circuit parameters and dispersion models against full-wave eigenmode simulations performed in COMSOL Multiphysics \cite{comsol}. Floquet-periodic boundary conditions were imposed on each pair of opposite unit-cell faces, with phase shifts prescribed by the corresponding Bloch-wavevector components.

The simulations used the unit-cell geometry and nominal material parameters specified in Fig.~\ref{Fig1-unitCell}. For the dispersion and isofrequency calculations, the conductors were modeled as perfect electric conductors and the dielectric was assumed lossless; accordingly, $R=0$. Any nonzero $\operatorname{Im}(k_x)$ therefore represents evanescence within a stopband rather than attenuation due to material absorption. Conductor and dielectric losses are included later in the finite-slab scattering analysis through the finite metal conductivity and dielectric loss tangent, respectively (Section~\ref{sect:s-parameters}).

Fig.~\ref{Fig4-dispersionRelation}(a) shows the dispersion relation for
$\Delta z=c/2=5~\mathrm{mm}$. In this symmetric configuration, $C_1=C_2=0.74~\mathrm{pF}$, yielding the maximum total series capacitance, $C=0.37~\mathrm{pF}$. The lower, dielectric-like passband extends from zero frequency to approximately $3.33~\mathrm{GHz}$, whereas the upper,
plasma-like passband begins at $f_{\mathrm{p}}=9.45~\mathrm{GHz}$.

Both analytical models reproduce the two passbands. The local-field model of Eqs.~\eqref{eq:dispersion_loadedWM}--\eqref{eq:spatialHarmonics} agrees more closely with the full-wave results because it retains the higher-order spatial harmonics of the lattice. Their omission in the multilayer model of Eqs.~\eqref{eq:disRel}--\eqref{eq:surfImpedance} produces larger deviations, particularly in the upper passband, where the unit cell is no longer deeply subwavelength. Nevertheless, the compact multilayer model enables closed-form estimates of the low-frequency refractive index and plasma frequency and facilitates the inclusion of losses in finite-slab scattering calculations.

Fig.~\ref{Fig4-dispersionRelation}(b) compares the isofrequency contours in the $xy$ plane obtained with COMSOL and the local-field model of Eqs.~\eqref{eq:dispersion_loadedWM}--\eqref{eq:spatialHarmonics}. Their agreement supports the representation of the SCCM as a capacitively loaded wire medium \cite{BelovS2002}. Within the equivalent-circuit approximation, the axial displacement primarily changes the series capacitance, whereas the unit-cell inductance remains approximately constant. For the distinct sliding split-strip geometry of Ref.~\cite{Enriquez2025p}, residual discrepancies in the dispersion were reduced using numerically retrieved circuit parameters. Here, we retain the geometry-derived RLC parameters to preserve the predictive character of the SCCM model.

\begin{table}[t]
\centering
\caption{In-plane anisotropy $\eta_{\mathrm{aniso}}(\%)=100(k_{\max}-k_{\min})/k_{\mathrm{mean}}$ of the SCCM isofrequency contours.}
\label{tab:anisotropy}
\footnotesize
\setlength{\tabcolsep}{3.5pt}
\begin{tabular}{c cc cc}
\hline
& \multicolumn{2}{c}{$\Delta z/c=0.2$}
& \multicolumn{2}{c}{$\Delta z/c=0.5$} \\
\cline{2-3}\cline{4-5}
$f$ (GHz) & Simulation & Theory & Simulation & Theory \\
\hline
$3.00$  & 0.01 & 0.01 & 0.86 & 0.70 \\
$12.5$ & 1.83 & 1.96 & 1.39 & 1.70 \\
\hline
\end{tabular}
\end{table}

The isofrequency contours reveal high in-plane isotropy near the $\Gamma$ point. To quantify this behavior, we define the percentage anisotropy coefficient as $\eta_{\mathrm{aniso}}(\%)=100(k_{\max}-k_{\min})/k_{\mathrm{mean}}$, where $k_{\max}$, $k_{\min}$, and $k_{\mathrm{mean}}$ are the maximum, minimum, and angular mean, respectively, of the radial Bloch wavenumber $k(\theta)=\sqrt{k_x^2+k_y^2}$ along a given contour. Thus, $\eta_{\mathrm{aniso}}=0\%$ corresponds to a perfectly circular contour. Table~\ref{tab:anisotropy} reports the simulated and analytical values at $3.00$ and $12.50~\mathrm{GHz}$, selected near the upper limits of the quasi-isotropic regions, for both $\Delta z/c=0.2$ and $0.5$. Although the contours for $\Delta z/c=0.2$ are omitted for brevity, Table~\ref{tab:anisotropy} shows that $\eta_{\mathrm{aniso}}$ remains below $2\%$ at both limits of the investigated tuning range and at both selected frequencies. The small values obtained from both approaches quantitatively confirm the weak directional dependence of the SCCM response. This behavior benefits directive antennas and gradient-index devices by providing a nearly direction-independent in-plane refractive index~\cite{Liu:09}. In the upper passband, the nearly isotropic response around the nonzero-frequency $\Gamma$-point cutoff may also reduce sensitivity to the in-plane propagation direction in mechanically tunable plasma haloscopes~\cite{millar2023alpha}.
\subsection{Low-frequency refractive index and plasma frequency}

The artificial-dielectric and artificial-plasma regimes are identified here from their Bloch-dispersion signatures. The low-frequency branch originates from the $\Gamma$ point at zero frequency and is approximately linear, $k_x\simeq n_0k_0$, where
\begin{equation}
n_0
=
\lim_{f\rightarrow0}\frac{k_x}{k_0}
\label{eq:n0Definition}
\end{equation}
is the low-frequency Bloch refractive index. This propagating branch is therefore interpreted as an artificial-dielectric response. By contrast, the upper branch emerges from the $\Gamma$ point at a nonzero frequency $f_{\mathrm{p}}$, defined by
\begin{equation}
k_x(f_{\mathrm{p}})=0,
\qquad
f_{\mathrm{p}}\neq0.
\label{eq:fpDefinition}
\end{equation}
The finite-frequency cutoff and absence of real Bloch wave vectors in the adjacent stopband are characteristic signatures of an artificial-plasma response, as shown in Fig.~\ref{Fig4-dispersionRelation}(a).

This interpretation is consistent with the quasistatic homogenization of reactively loaded wire media presented in Ref.~\cite{BelovS2002}. For propagation transverse to the wire axis, that model yields a Lorentz-type effective permittivity for capacitively and series-LC-loaded wires. Its positive low-frequency limit describes the artificial-dielectric regime, whereas its zero crossing above the circuit resonance defines the plasma frequency and the onset of the plasma-like passband. Because the SCCM is spatially dispersive, this quasistatic model is used only to support the physical interpretation; the finite-wavevector response is described using the Bloch dispersion relations and isofrequency contours. The following analysis therefore focuses on the geometrical tunability of $n_0$ and $f_{\mathrm{p}}$.

In the low-frequency limit, the sheet impedance is dominated by its capacitive contribution:
\begin{equation}
Z_s
\simeq
\frac{b}{c}\frac{1}{\jj\omega C}.
\label{eq:lowFrequencySheetImpedance}
\end{equation}
Substituting Eq.~\eqref{eq:lowFrequencySheetImpedance} into Eq.~\eqref{eq:disRel} and consistently expanding the complete dispersion relation to second order in $k_xa$ and $k_0a$ gives $k_x\simeq n_0k_0$, with
\begin{equation}
n_0
\approx
\sqrt{
1+\frac{C}{\varepsilon_0ab/c}
}
=
\sqrt{
1+\frac{cC}{\varepsilon_0ab}
}.
\label{eq:refIndA}
\end{equation}

The plasma frequency is the nonzero-frequency solution of $k_x=0$. Substituting this condition into Eq.~\eqref{eq:disRel} and using $(1-\cos x)/\sin x=\tan(x/2)$ gives
\begin{equation}
\tan\left(\frac{k_{\mathrm{p}}a}{2}\right)
=
\jj
\frac{Z_0c}
{2b\left(
R+\jj\omega_{\mathrm{p}}L
+\dfrac{1}{\jj\omega_{\mathrm{p}}C}
\right)},
\label{eq:PlasmaFreqCondition}
\end{equation}
where
\begin{equation}
\omega_{\mathrm{p}}=2\pi f_{\mathrm{p}},
\qquad
k_{\mathrm{p}}=\frac{\omega_{\mathrm{p}}}{c_0},
\end{equation}
$c_0$ is the speed of light in vacuum, and $L$ is the total unit-cell inductance given by Eq.~\eqref{eq:inductance}. Equation~\eqref{eq:PlasmaFreqCondition} is transcendental and can be solved numerically within the multilayer model without introducing any further small-$k_{\mathrm{p}}a$ expansion.

To obtain a closed-form estimate, we set $x=k_{\mathrm{p}}a$ and use the rational approximation
\begin{equation}
\tan\left(\frac{x}{2}\right)
\approx
\frac{\pi^2x}{2\pi^2-2x^2}.
\label{eq:tangentApproximation}
\end{equation}
This approximation reproduces the poles at $x=\pm\pi$ and reduces to $x/2$ near the origin. Assuming low losses and setting $R=0$, Eq.~\eqref{eq:PlasmaFreqCondition} then reduces to an algebraic equation for $\omega_{\mathrm{p}}^2$, yielding
\begin{equation}
\omega_{\mathrm{p}}
\approx
\sqrt{
\frac{
1+\dfrac{\varepsilon_0ab}{cC}
}{
L\dfrac{\varepsilon_0ab}{c}
+\dfrac{\mu_0\varepsilon_0a^2}{\pi^2}
}
}.
\label{eq:plasmaFreqA}
\end{equation}
Equations~\eqref{eq:refIndA} and~\eqref{eq:plasmaFreqA} explicitly relate the tunable capacitance to the two principal electromagnetic characteristics of the SCCM. Increasing $C$ increases the low-frequency refractive index $n_0$ while decreasing the plasma frequency $f_{\mathrm{p}}$. Because $C$ is controlled by the relative axial displacement of the conductors, these relations provide a direct analytical description of the geometrical tunability of the artificial-dielectric and artificial-plasma responses.

\section{Numerical Analysis}

\subsection{Validation of Unit-Cell Fields and Currents}
\begin{figure*}
\centering
\includegraphics[width=0.9\textwidth]{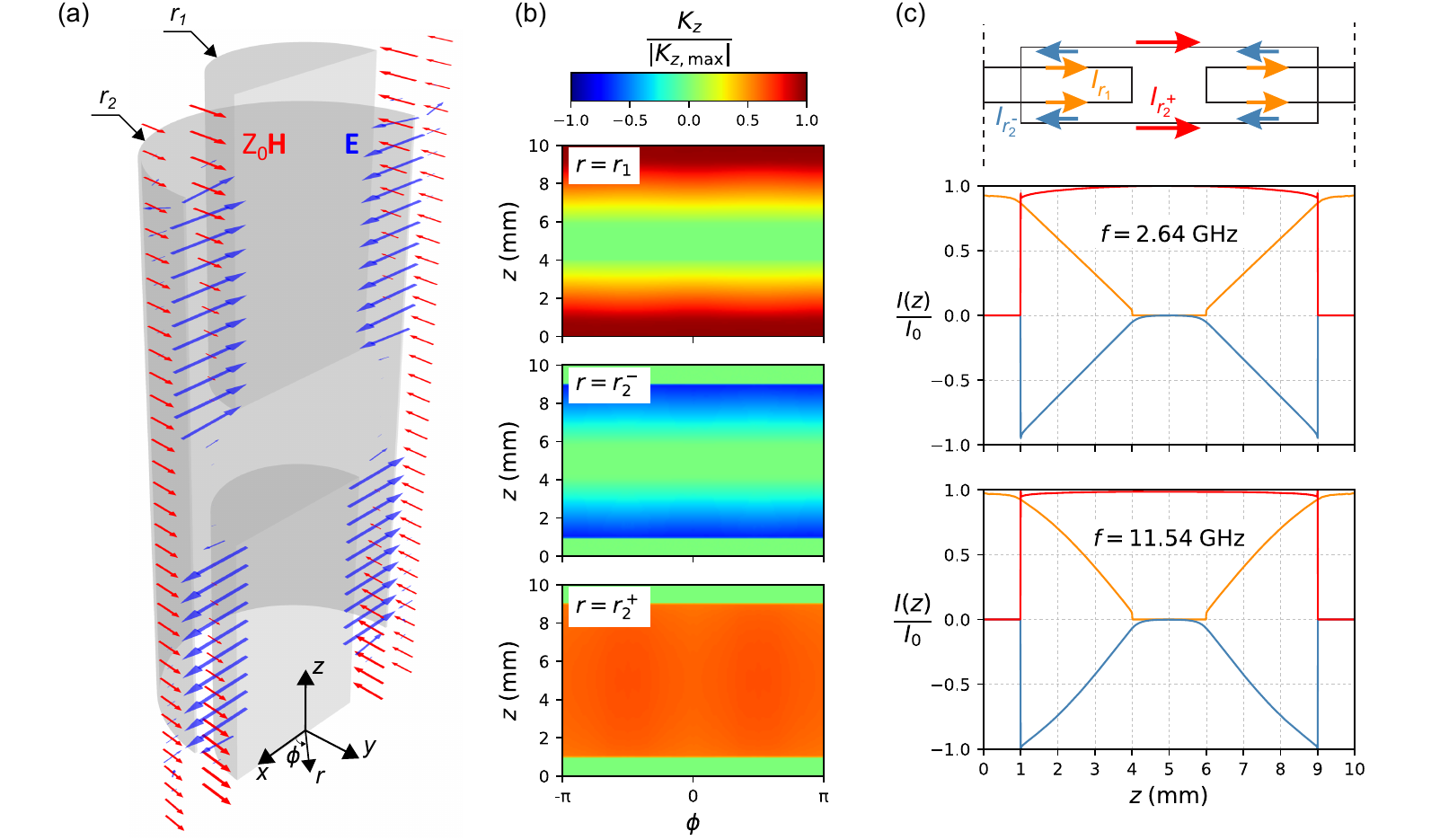}
  \caption{Simulated fields and currents in the SCCM unit cell of Fig.~\ref{Fig1-unitCell} under Floquet-periodic boundary conditions, for a Bloch mode with $k_xa=\varphi_x=90^\circ$ and $\varphi_y=\varphi_z=0$. (a) Electric-field (blue) and magnetic-field (red) vectors near the conductor surfaces at $f=2.64~\mathrm{GHz}$. (b) Normalized axial surface-current density $K_z/|K_{z,\max}|$ on the inner-conductor surface ($r=r_1$) and the inner and outer surfaces of the outer conductor ($r=r_2^-$ and $r=r_2^+$, respectively). (c) Normalized axial current profiles $I(z)/I_0$ at $f=2.64~\mathrm{GHz}$ (upper, artificial-dielectric passband) and $f=11.54~\mathrm{GHz}$ (lower, artificial-plasma passband).}
 \label{Fig5-fieldMonitors}
\end{figure*}

The derivation of the equivalent RLC parameters relies on the assumption that the surface current is independent of the azimuthal coordinate and varies piecewise along the cylinder axes, as illustrated in Fig.~\ref{Fig2-currentApprox}. To assess the validity of this assumption, we performed eigenfrequency simulations of the three-dimensional periodic unit cell shown in Fig.~\ref{Fig1-unitCell} using COMSOL Multiphysics~\cite{comsol} with Floquet-periodic boundary conditions. The geometrical parameters are the same as those introduced in Section~\ref{section2:theory}.

We first examine the electromagnetic-field distributions. Fig.~\ref{Fig5-fieldMonitors}(a) shows representative electric-field vectors (blue arrows) and magnetic-field vectors (red arrows) at $k_xa=90^\circ$, corresponding to the midpoint of the $\Gamma$--$X$ path along the first dispersion branch and a frequency of $f=2.64~\mathrm{GHz}$. As expected, the electric field is concentrated primarily in the coaxial overlap regions that form the cylindrical capacitors. By contrast, the magnetic field extends through both the internal and external regions of the unit cell and remains comparatively uniform outside the outer conductor. This spatial separation is potentially attractive for magnetic resonance imaging (MRI), where coaxially shielded metamaterial resonators have been employed to confine the electric field and reduce its interaction with surrounding tissues while maintaining a strong externally accessible magnetic field~\cite{Zhu2024Wearable,Wu2024Wireless}. Related coaxially shielded textile metamaterials have also been applied to near-field body-area communication networks~\cite{Zhu2024BodyArea}. In the SCCM, an analogous field separation is obtained for the TM Bloch mode considered here.

Fig.~\ref{Fig5-fieldMonitors}(b) shows the surface-current density sampled along circumferential paths on the conductor surfaces, confirming its approximate invariance with respect to the azimuthal coordinate. Fig.~\ref{Fig5-fieldMonitors}(c) presents the corresponding axial current profiles for Bloch waves propagating along the $x$ direction at $k_xa=\pi/2$. The upper and lower panels of Fig.~\ref{Fig5-fieldMonitors}(c) correspond to the first artificial-dielectric passband at $f=2.64~\mathrm{GHz}$ and the second artificial-plasma passband at $f=11.54~\mathrm{GHz}$, respectively. Although not shown here, simulations performed for other axial displacements and Bloch propagation constants exhibit similar current distributions. On the inner conductor, the current reaches its maximum near the outer-conductor gap at the unit-cell boundary and decreases toward zero at the edges of the inner-conductor gap. Within the overlapping region, the current induced on the inner surface of the outer conductor mirrors that on the inner conductor but flows in the opposite direction. By contrast, the current on the exterior surface of the outer conductor remains nearly uniform along the $z$ direction. Deviations from the piecewise-linear current distribution assumed in Fig.~\ref{Fig2-currentApprox} become more pronounced in the second passband. These deviations likely contribute to the slightly reduced agreement between the analytical models and the full-wave numerical results at higher frequencies.
\subsection{Validation of the Scattering Parameters} \label{sect:s-parameters}


\begin{figure}
  \centering
  \includegraphics[width=0.9\columnwidth]{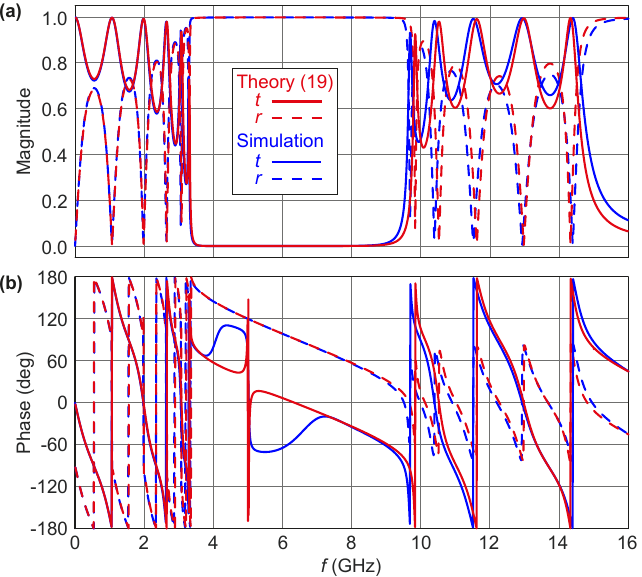}
  \caption{Theoretical and simulated transmission and reflection under normal incidence, magnitude (a) and phase (b), through a slab of thickness $6a$ (six periods stacked along $x$). The shift applied to the inner conductor of the coaxial cable is $\Delta z = 5$~mm.}
  \label{Fig6-transmissionReflection}
\end{figure}
From the dispersion relation~\eqref{eq:disRel} and normalized admittance~\eqref{eq:admittance}, we obtain the Bloch propagation constant $k_x$ and modal impedance $Z$, respectively. We then calculate the complex transmission and reflection coefficients of an $N$-layer SCCM slab embedded in free space as~\cite{Parra2025,Zhuravlev2026}
\begin{align}
t &=\frac{2Z_0Z}{2Z_0Z\cos(k_xNa)
  +\jj\left(Z^2+Z_0^2\right)\sin(k_xNa)}, \nonumber\\
r &=\frac{\jj\sin(k_xNa)\left(Z^2-Z_0^2\right)}
{2Z_0Z\cos(k_xNa)
  +\jj\left(Z^2+Z_0^2\right)\sin(k_xNa)}.
\label{eq:trns_thr_MM}
\end{align}
where $N$ is the number of layers.

We validate these expressions for a six-layer slab ($N=6$) by comparison with full-wave CST simulations \cite{cst}. The simulation employed unit-cell boundary conditions along the $y$ and $z$ directions and Floquet ports on the boundaries normal to the propagation direction. It is worth noting that the port reference planes were placed half a period away from the center of the peripheral split coaxial cables. Both models use the unit-cell parameters specified in Section~\ref{section2:theory} and include a dielectric loss tangent of $\tan\delta=10^{-3}$ and a metal conductivity of $\sigma=5.96\times10^{7}~\mathrm{S/m}$. In the analytical model, the dielectric and conductor losses enter the surface impedance through the complex capacitance and resistance $R$, respectively. 

Fig.~\ref{Fig6-transmissionReflection}(a) and (b) compare the magnitude and phase, respectively, of the calculated and simulated transmission and reflection coefficients. The model reproduces the low-pass response of the artificial-dielectric band, including the small Fabry--Pérot ripples caused by the impedance mismatch between free space and the finite slab. The transmission-phase curves differ only where $|t|$ is strongly suppressed and $|\operatorname{Im}(k_x)|$ is largest. In these regions, the phase of a complex quantity with nearly zero magnitude is ill-defined and highly sensitive to numerical noise, explaining the observed discrepancy. Elsewhere, the close agreement confirms the accuracy of the model for finite SCCM slabs.



\subsection{Tunability of the Low-Frequency Refractive Index and Plasma Frequency}
\begin{figure}
  \centering
  \includegraphics[width=0.9\columnwidth]{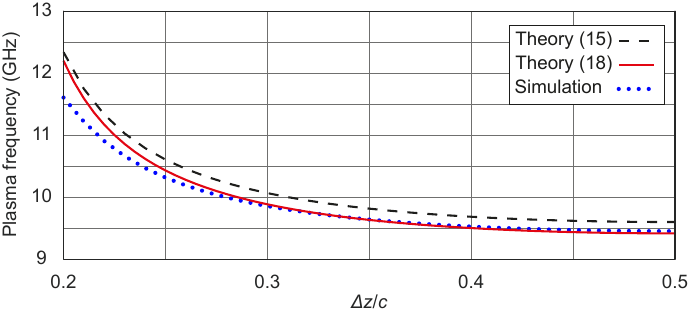}
  \caption{Plasma frequency of the SCCM as a function of the normalized axial displacement $\Delta z/c$ of the inner conductor.}
  \label{Fig7-plasmaFreq}
\end{figure}

\begin{figure}
  \centering
  \includegraphics[width=0.9\columnwidth]{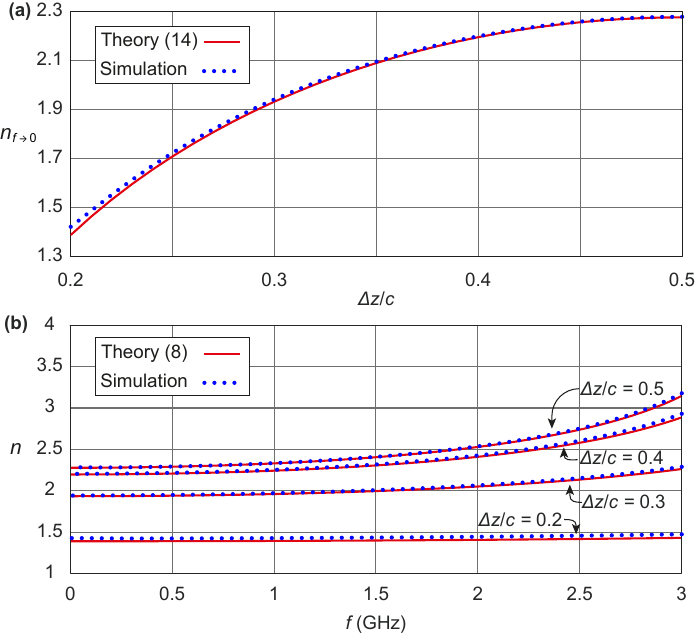}
  \caption{(a) Low-frequency Bloch refractive index tunability of the SCCM as a function of the normalized axial displacement $\Delta z/c$ of the inner conductor. (b) Frequency dispersion of the refractive index for  different values of $\Delta z/c$.}
   \label{Fig8-refractiveIndex}
\end{figure}

We investigate the tunability of the SCCM using the CST Eigenmode Solver, enclosing the unit cell with periodic boundary conditions in all directions, for axial displacements ranging from $\Delta z=2$ to $5~\mathrm{mm}$. For each displacement, the simulated low-frequency refractive index is extracted by fitting the artificial-dielectric branch near the $\Gamma$ point to $k_x\simeq n_0k_0$, whereas the simulated plasma frequency is obtained from the nonzero-frequency intersection of the artificial-plasma branch with $k_x=0$. The corresponding analytical values of $n_0$ are calculated using Eq.~\eqref{eq:refIndA}. The plasma frequency is obtained from the closed-form approximation in Eq.~\eqref{eq:plasmaFreqA}. Figs.~\ref{Fig7-plasmaFreq} and~\ref{Fig8-refractiveIndex}(a) show the resulting dependence of $f_{\mathrm{p}}$ and $n_0$ on $\Delta z$, respectively.

Fig.~\ref{Fig7-plasmaFreq} shows that $f_{\mathrm{p}}$ decreases as the two overlap lengths become more balanced and the equivalent series capacitance increases. To quantify the achievable variation, we define the tunability of a parameter $X$ as $T_X=2(X_{\max}-X_{\min})/(X_{\max}+X_{\min})\times100\%$, where $X_{\max}$ and $X_{\min}$ are the maximum and minimum values, respectively. The closed-form Eq. \eqref{eq:plasmaFreqA} predicts a plasma-frequency tunability of approximately $26\%$, from $f_{\mathrm{p,min}}=9.42~\mathrm{GHz}$ to $f_{\mathrm{p,max}}=12.20~\mathrm{GHz}$, whereas the full-wave simulations yield approximately $21\%$, from $f_{\mathrm{p,min}}=9.45~\mathrm{GHz}$ to $f_{\mathrm{p,max}}=11.61~\mathrm{GHz}$. Although larger tuning
ranges have been reported, those approaches typically rely on transverse wire displacements~\cite{Kowitt2023,sakhno2025volume,Lindahl2026}. By contrast, the SCCM tunes $f_{\mathrm{p}}$ through axial motion, changing the capacitive overlap while preserving the transverse lattice geometry. This mechanism may be advantageous for mechanically tunable wire-medium haloscopes.

Fig.~\ref{Fig8-refractiveIndex}(a) shows that the analytical and simulated values of $n_0$ follow the same displacement dependence. The analytical model predicts a low-frequency refractive-index tunability of approximately $49\%$, from $n_{\mathrm{0,min}}=1.39$ to $n_{\mathrm{0,max}}=2.28$,  whereas the full-wave simulations
yield approximately $46\%$, from $n_{\mathrm{0,min}}=1.42$ to $n_{\mathrm{0,max}}=2.28$. Fig.~\ref{Fig8-refractiveIndex}(b) further shows that the frequency-dependent Bloch index, $n(f)=k_x/k_0$, remains approximately constant up to
$2~\mathrm{GHz}$ throughout the tuning range. This weakly dispersive response is attractive for tunable GRIN devices because a frequency-stable refractive-index profile can reduce chromatic aberration while enabling microwave focusing and beam shaping~\cite{Smith2005}.


\section{Conclusion}
The results establish the relative axial displacement $\Delta z$ as an effective mechanism for continuously tuning the SCCM dispersion in both its artificial-dielectric and plasma-like regimes. Varying $\Delta z$ changes the total series capacitance without altering the transverse lattice geometry, while the unit-cell inductance remains approximately constant. Consequently, $\Delta z$ provides a single mechanical control parameter for the tuning of the low-frequency Bloch refractive index and plasma frequency. The local-field model closely reproduces the simulated dispersion and isofrequency contours, confirming the importance of higher-order spatial harmonics, particularly in the upper passband. Although less accurate when the unit cell is not deeply subwavelength, the multilayer homogenization model captures the principal dispersion trends, yields the modal impedance and closed-form estimates of the low-frequency refractive index and plasma frequency, and enables loss-inclusive finite-slab scattering calculations. Full-wave simulations yield tunabilities of $46\%$ in the low-frequency Bloch refractive index and $21\%$ in the plasma frequency. The nearly circular isofrequency contours further confirm high in-plane isotropy near the $\Gamma$ point in both passbands. The combination of this tunability and isotropy makes the SCCM promising for reconfigurable GRIN devices, directive antennas, and mechanically tunable plasma haloscopes.

\section*{CRediT authorship contribution statement}

\textbf{Alexander Zhuravlev:} Methodology, Software, Formal analysis, Investigation, Visualization, Writing -- original draft.
\textbf{Jim A. Enriquez:} Methodology, Software, Formal analysis, Investigation, Visualization, Writing -- original draft, Writing -- review \& editing.
\textbf{Pavel A. Belov:} Supervision, Funding acquisition, Writing -- review \& editing.
\textbf{Juan D. Baena:} Conceptualization, Methodology, Formal analysis, Visualization, Supervision, Project administration, Writing -- original draft, Writing -- review \& editing.

\section*{Declaration of Competing Interest}

The authors declare that they have no known competing financial interests
or personal relationships that could have appeared to influence the work
reported in this paper.


\section*{Funding}
The theoretical analysis of this work was supported by the Ministry of Science and Higher Education of the Russian Federation under Project FSER-2025-0009. The numerical simulations were funded by the Russian Science Foundation, Grant No. 25-12-00261
(https://rscf.ru/project/25-12-00261/).

\section*{Acknowledgments}
The authors thank Denis Sakhno for providing preliminary benchmark calculations used to verify the isofrequency-contour implementation.

\section*{Data Availability}

The data that support the findings of this study are available from the
corresponding author upon reasonable request.
\bibliographystyle{cas-model2-names}
\bibliography{references}

\end{document}